%% file: letter_v3.tex
\documentclass[aps,prl,twocolumn,superscriptaddress,preprintnumbers,longbibliography,nofootinbib]{revtex4-1}
\usepackage{amsmath}
\usepackage{amssymb}
\usepackage{graphicx}
\usepackage{gensymb}
\usepackage{float}
\usepackage{physics}
\usepackage{comment}
\usepackage[colorlinks = true, citecolor = blue, linkcolor = blue, urlcolor = blue]{hyperref}
\def\bea#1\eea{\begin{align}#1\end{align}}

\newcommand{\bef}{\begin{figure}[h!tb]\centering}
\newcommand{\eef}{\end{figure}}
\allowdisplaybreaks

\input{packages}

\input{definitions}

\begin{document}

\title{Determination of the strong coupling constant and the Collins--Soper kernel \\
from the energy--energy correlator in $e^+e^-$ collisions}

\author{Zhong-Bo Kang}
\email{zkang@physics.ucla.edu}
\affiliation{Department of Physics and Astronomy, University of California, Los Angeles, CA 90095, USA}
\affiliation{Mani L. Bhaumik Institute for Theoretical Physics, University of California, Los Angeles, CA 90095, USA}
\affiliation{Center for Frontiers in Nuclear Science, Stony Brook University, Stony Brook, NY 11794, USA}

\author{Jani Penttala}
\email{janipenttala@physics.ucla.edu}
\affiliation{Department of Physics and Astronomy, University of California, Los Angeles, CA 90095, USA}
\affiliation{Mani L. Bhaumik Institute for Theoretical Physics, University of California, Los Angeles, CA 90095, USA}

\author{Congyue Zhang}
\email{maxzhang2002@g.ucla.edu}
\affiliation{Department of Physics and Astronomy, University of California, Los Angeles, CA 90095, USA}
\affiliation{Mani L. Bhaumik Institute for Theoretical Physics, University of California, Los Angeles, CA 90095, USA}

\begin{abstract}
We have conducted the first simultaneous global fit of the strong coupling constant $\alpha_s$ and the Collins--Soper (CS) kernel using the  energy--energy correlators (EEC) of $e^{+} e^{-}$ collisions in the back-to-back limit. This analysis, based on the transverse-momentum-dependent (TMD) factorization of EEC at next-to-next-to-next-to-leading logarithmic ($\rm{N}^{3} \rm{LL}$) accuracy, yields $\alpha_s$ consistent with the world average. We have tested two different parametrizations for the non-perturbative CS kernel and found both  to align with results obtained from lattice QCD and fits on semi-inclusive deep inelastic scattering and Drell--Yan process.
\end{abstract}

\maketitle 

{\it \textbf{Introduction.}}
One of the most important parameters in quantum chromodynamics (QCD) is the strong coupling constant $\as$, which describes the strength of strong interactions.
Its precise value depends on the relevant momentum scale of the underlying process, and this momentum-scale dependence can be calculated perturbatively using a renormalization group equation~\cite{Gross:1973id,Politzer:1973fx,Baikov:2016tgj,Luthe:2016ima,Herzog:2017ohr,Luthe:2017ttg,Chetyrkin:2017bjc}.
This equation requires, however, an initial condition that has to be fitted to the experimental data, and usually this initial condition
is chosen as the value of the coupling constant at the $Z$-boson mass, $\as(m_Z)$.
This value of the strong coupling constant can then be used to describe any QCD interaction, provided that the momentum scale is large enough for the process to be perturbative.
 
The strong coupling constant has been previously extracted from a plethora of different processes~\cite{ParticleDataGroup:2024cfk}, demonstrating the perturbative consistency of QCD.
In terms of determining the precise value of the coupling constant, an especially powerful class of processes  has turned out to be event-shape observables~\cite{Abbate:2010vw,Hoang:2015hka,H1:2017bml} where, instead of studying the individual produced particles, one examines the overall geometry of the final-state distribution.
This means that event-shape observables are generally less sensitive to the non-perturbative (NP) hadronization of the final-state particles, allowing one to focus more on other parts of the collision process.

The process of interest in this Letter is the energy--energy correlator (EEC) event-shape observable which describes angular correlations between pairs of produced particles.
EEC was one of the first infrared-safe observables for QCD~\cite{Basham:1978bw,Basham:1978zq}, and it has already been studied extensively at different experiments~\cite{SLD:1994idb, L3:1992btq, OPAL:1991uui, TOPAZ:1989yod, TASSO:1987mcs, JADE:1984taa, Fernandez:1984db, Wood:1987uf, CELLO:1982rca, PLUTO:1985yzc, OPAL:1990reb, ALEPH:1990vew, L3:1991qlf, SLD:1994yoe,CMS:2024mlf, ALICE:2024dfl}.
Recently, it has garnered renewed interest in precision QCD research, probing both perturbative and non-perturbative dynamics in collisions ranging from $e^+e^-$ annihilation and deep inelastic $ep$ scattering to proton--proton and heavy-ion collisions~\cite{Neill:2022lqx}. Two kinematic limits of EEC have been extensively studied and have shown important physics insights of its own~\cite{Liu:2024lxy,Ebert:2020sfi}: the collinear limit~\cite{Dixon:2019uzg}, where a pair of particles moves in the same or nearly collinear direction, and the back-to-back limit, which focuses on particle pairs with momenta pointing in opposite directions.

In this Letter, we focus on the back-to-back limit of EEC. This limit can be described efficiently using the framework of transverse-momentum-dependent (TMD) factorization~\cite{Collins:2011zzd,Boussarie:2023izj}, which factorizes the collision into separate calculable parts of different momentum scales. To account for non-perturbative hadronization effects, we introduce the NP Sudakov factor and the NP Collins--Soper (CS) kernel, also known as the rapidity anomalous dimension, which is universal across different processes~\cite{Collins:2011zzd,Boussarie:2023izj,Shu:2023cot}. A precise determination of the CS kernel is of especially high interest due to its crucial role in understanding the rapidity evolution of TMD parton distributions functions (PDF) and fragmentation functions (FF),
and it has emerged as a central focus in recent lattice QCD calculations~\cite{Shanahan:2020zxr,LatticeParton:2020uhz,LatticePartonLPC:2022eev,Shu:2023cot,LatticePartonLPC:2023pdv,Avkhadiev:2023poz,Avkhadiev:2024mgd,Bollweg:2024zet}. Understanding the NP CS kernel is thus crucial for any study based on TMD factorization, especially for imaging nuclear structure in the transverse momentum plane---one of the major science goals of the future Electron--Ion Collider~\cite{Accardi:2012qut,Aschenauer:2017jsk,AbdulKhalek:2021gbh}.

In this Letter, we present a simultaneous fit of both the strong coupling constant and the Collins--Soper kernel at next-to-next-to-next-to-leading logarithmic (N$^3$LL) accuracy to the EEC data from $e^+ e^-$ collisions. This marks the first extraction of the CS kernel from an observable that is independent of non-perturbative parton distribution and fragmentation functions, thereby minimizing reliance on other non-perturbative components in the calculation. This highlights the EEC as a particularly clean observable for precision studies of QCD. Our fit for the non-perturbative part of the CS kernel can be applied to other processes described using TMD factorization, allowing for more accurate analyses of TMD parton distributions and fragmentation functions.

{\it\textbf{Theoretical Formalism.}}
The EEC observable is defined as an event shape to measure the energy-weighted angular distance between pairs of particles \cite{Ebert:2020sfi}:
\begin{equation}
  \dv{\Sigma}{\chi} = \sum_{ij} \int \dd{\sigma} \frac{E_i E_j}{Q^2} \delta\qty(\chi - \theta_{ij} ).
\end{equation}
Here $E_i$ is the energy of the particle $i$ and $Q$ is the center-of-mass energy of the collision. The sum goes over all final-state particles, and $\theta_{ij}$ is the angle between the the final-state particles $i$ and $j$. EEC is also commonly defined in the variable $z=(1-\cos \chi)/2$. Higher-order calculations of the EEC involve singular contributions of the form $\log^n (1-z)$ that must be resummed in the back-to-back limit where $z \to 1$, whereas the non-singular contributions remain unchanged. We can write the EEC as
\begin{align}
\frac{\dd{\Sigma}}{\dd{\chi}} = \frac{\sin (\chi)}{2} \frac{\dd{\Sigma}}{\dd{z}},
\quad
\frac{\dd{\Sigma}}{\dd{z}} = \frac{\dd{\Sigma^{\rm{res}}}}{\dd{z}} + 
\frac{\dd{\Sigma^{\rm{nons}}}}{\dd{z}}\,,
\end{align}
where $\sigma^{\rm{res}}$ and $\sigma^{\rm{nons}}$ correspond to the resummed singular and non-singular contributions to the EEC.
The non-singular contribution has been computed analytically at NLO in~\cite{Dixon:2018qgp}, and the fixed-order EEC has been obtained numerically at NNLO~\cite{Tulipant:2017ybb}. In this Letter, we use the NLO result.
The resummation of the large logarithms in the back-to-back limit can be done using the TMD factorization and leads to the expression~\cite{Ebert:2020sfi}
\begin{align}
&\frac{\dd{\Sigma^{\rm{res}}}}{\dd{z}} = \frac{\hat{\sigma}_0}{8} H_{q\bar{q}}(Q, \mu^i_H) \int_0^\infty \dd{(b Q)}^2 J_0\qty(b Q \sqrt{1-z}) 
\\
&\times J^{\rm{pert}}_q(b_*, \mu^i_J, \zeta^i) J^{\rm{pert}}_{\bar{q}} (b_*, \mu^i_J, \zeta^i) \left(\frac{\zeta^f}{\zeta^i}\right)^{K(b,\mu^f)} e^{-2 S_{\rm{NP}}(b)}
\nonumber\\ 
&\times \exp{\int^{\mu^f}_{\mu_H^i} \frac{\dd{\mu'}}{\mu'} \gamma_{H}(Q,\mu') + 
2 \int^{\mu^f}_{\mu_J^i} \frac{\dd{\mu'}}{\mu'} \gamma_{J}(\mu',\zeta^i)}.  
\nonumber
\end{align}
Here $\hat{\sigma}_0$ is the Born cross section of $e^+ + e^- \rightarrow q + \bar{q}$, $H_{q\bar{q}}$ is the hard function, and $J_q$ ($J_{\bar q}$) is the quark (anti-quark) jet function. The hard function and the jet functions satisfy the renormalization group equations for the momentum scale $\mu$, which is given by
the hard anomalous dimension $\gamma_{H}$ and jet anomalous dimension $\gamma_{J}$, respectively.
The jet functions also depend on the rapidity scale $\zeta$ through the CS evolution that is governed by the CS kernel $K$.
Our resummation is performed at the $\rm{N}^3\rm{LL}$ accuracy and relies on the 2-loop  hard \cite{Becher:2008cf}, jet \cite{Ebert:2020sfi}, and soft \cite{Li:2016ctv} functions, the 3-loop non-cusp anomalous dimensions \cite{Ebert:2020yqt,Luo:2019szz,Li:2016ctv} and rapidity anomalous dimension \cite{Lubbert:2016rku,Vladimirov:2016dll,Li:2016ctv}, and the 4-loop beta function 
\cite{Tarasov:1980au,Larin:1993tp,vanRitbergen:1997va,Czakon:2004bu} and cusp anomalous dimension $\Gamma_{\rm{cusp}} $ \cite{Korchemsky:1987wg,Henn:2019swt,Moch:2004pa,Vogt:2004mw,Henn:2016wlm,Moch:2017uml,Lee:2019zop,Henn:2019rmi,Bruser:2019auj,vonManteuffel:2020vjv}.
We have implemented the standard $b_*$ prescription~\cite{Collins:1984kg,Collins:2014jpa} to avoid the Landau pole of QCD, where $b_*=b/\sqrt{1+b^2/b_{\rm max}^2}$, and we choose $b_{\rm max} = 2e^{-\gamma_E} \ \rm{GeV}^{-1}$.
The CS kernel is defined as
\begin{align}
K(b,\mu) = &\, -2 \int_{\mu_{b_*}}^{\mu} \frac{d \mu'}{\mu'} \Gamma_{\rm{cusp}}[\alpha_s(\mu')]  
\nonumber\\
&\,-2 D_{\rm{pert}}(b_*,\mu_{b_*})  - 2  D_{\rm{NP}}(b)\,,
\label{eq:K}
\end{align}
where we also include the non-perturbative contribution $D_{\rm{NP}}$ that has to be modeled.
The perturbative boundary term $D_{\rm{pert}}$ of the CS kernel can be found in \cite{Li:2016ctv}. 
The canonical initial scales which eliminate logarithmic terms in the hard and jet functions are given by
\begin{align}
\mu_H^i &= Q, 
&
\mu_J^i &= \mu_{b_*}, 
&
\zeta^i &= \mu_{b_*}^2,
\end{align}
where $\mu_{b_*} = 2e^{-\gamma_E}/b_*$. For the final scales, we set $\mu^f=Q$ and $\zeta^f = Q^2$. 

There are two main sources of NP contributions. Firstly, the jet function $J_q$ is related to the TMD fragmentation function as follows: $J_q(b, \mu, \zeta) = \sum_{h}\int_0^1 \dd{z} \, z \, D_{h/q}(z, b, \mu, \zeta)$. Thus, when $1/b \lesssim \Lambda_{\rm QCD}$, the jet function receives a NP contribution, and we incorporate it within the NP Sudakov factor: $J_q = J_q^{\rm pert} \times e^{-S_{\rm{NP}}(b)}$, where $J_q^{\rm pert}$ is the perturbative result~\cite{Ebert:2020sfi} and for $S_{\rm{NP}}(b)$ we use the following form from~\cite{Kang:2024otf}:
\begin{equation}
S_{\rm{NP}} (b) = a_1 b^{a_2}.
\end{equation}
The second NP source arises from the NP contribution $D_{\rm NP}(b)$ of the CS kernel $K$ in Eq~\eqref{eq:K}, and we inspect two different parametrizations:
\begin{align}
D^{\rm{Fit\, 1}}_{\rm{NP}} = g_2 \,b\,  b_*
, \quad
D^{\rm{Fit\, 2}}_{\rm{NP}} = g'_2 \,\ln(\frac{b}{b_*}).
\end{align}
Here $a_1$ and $a_2$ in the NP Sudakov factor and $g_2$ (or $g_2'$) in the NP CS kernel are fit parameters that will be determined by the data.
The behavior of Fit 1 was motivated in~\cite{Vladimirov:2020umg} and applied in~\cite{Scimemi:2019cmh}, albeit with a different choice of $b_{\rm max}$. The parametrization for Fit 2 has been employed in~\cite{Sun:2014dqm,Echevarria:2020hpy}. Both extracted NP CS kernels have been extensively used in phenomenological TMD studies.

\begin{table}[t!]
\centering
\begin{tabular}{c c c c c}
\hline
\hline
Collaboration & $Q$\,(\rm{GeV}) & $\mathrm{N}_\mathrm{data}$ & $\chi^2 \,\mathrm{(Fit\,1)}$ & $\chi^2 \,\mathrm{(Fit\,2)}$ \\
\hline
OPAL~\cite{OPAL:1991uui} & 91.2 & 30 & 12.4 & 12.4\\
SLD~\cite{SLD:1994idb}    & 91.2 & 9 & 2.5 & 2.4\\
TOPAZ~\cite{TOPAZ:1989yod}   & 59.5 & 9 & 11.3 & 11.3\\
TOPAZ~\cite{TOPAZ:1989yod}  & 53.3 & 9 & 12.7 & 12.7\\
TASSO~\cite{TASSO:1987mcs}  & 43.5 & 9 & 8.6 & 8.7\\
TASSO~\cite{TASSO:1987mcs}  & 34.8 & 9 & 10.3 & 10.4\\
MARKII~\cite{Wood:1987uf} & 29.0 & 9 & 12.8 & 12.8\\
MAC~\cite{Fernandez:1984db}    & 29.0 & 9 & 14.2 & 14.3\\
Total  &      & 93 & 84.8 & 84.9\\
\hline
\hline
\end{tabular}
\caption{The experimental data used in the fits and the corresponding $\chi^2$ values for the best fits.}
\label{table1}
\end{table}

\begin{figure*}[t!]
    \centering
    \includegraphics[width =1\textwidth]{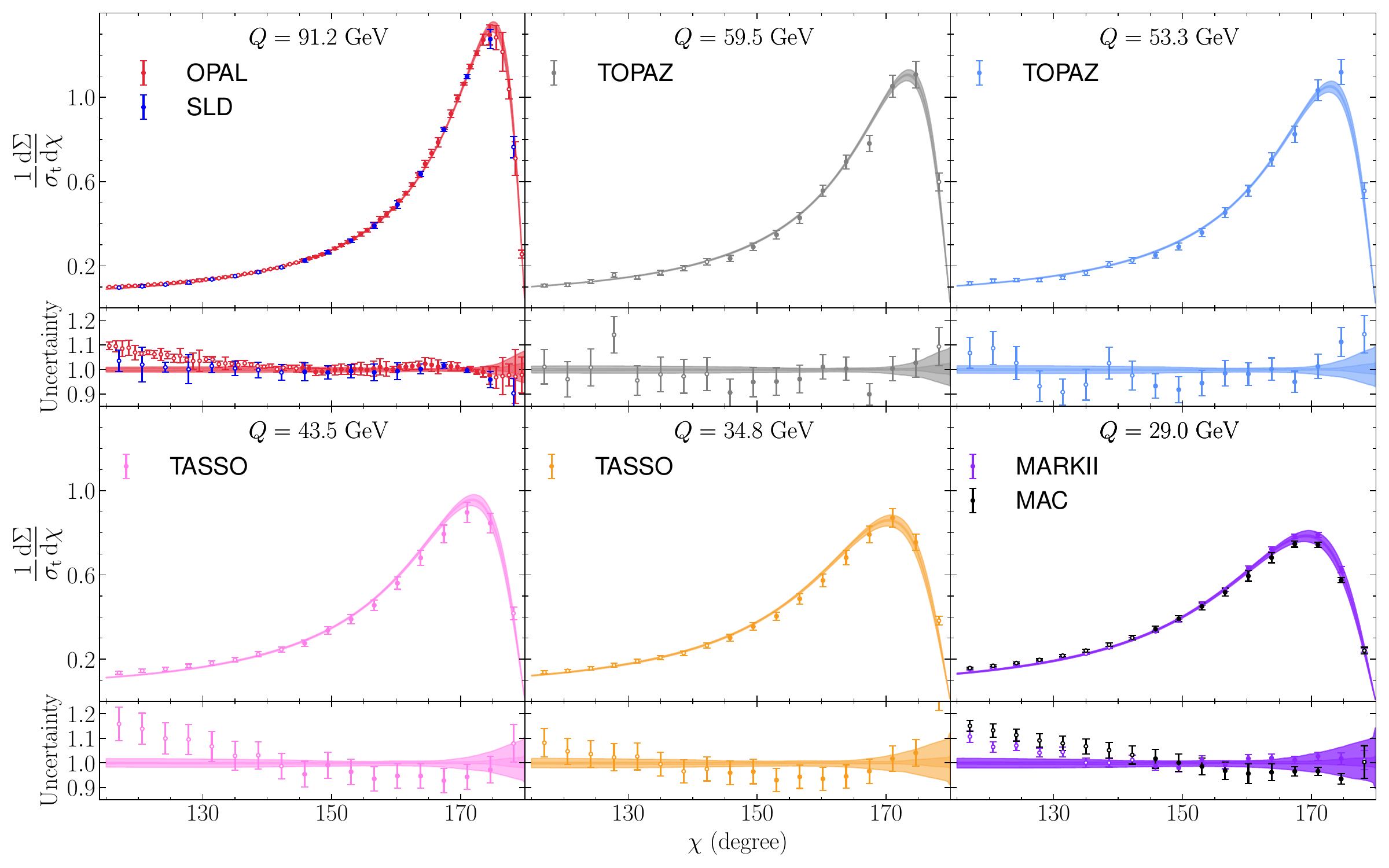} 
    \caption{Comparison of Fit 1's prediction with central parameters to the experimental data. The dark uncertainty band represents the middle 68\% fit uncertainty, and the light band represents the middle 68\% theoretical uncertainty. The solid points represent fitted data while empty points are not included in the fit. The total cross-section $\sigma_{\rm{t}}$ is calculated at 2-loop~\cite{Kardos:2018kqj}.}
    \label{FIG1}
\end{figure*}

{\it \textbf{Data Selection.}} 
Within the $Q$ range of $29.0$ GeV to $91.2$ GeV, we include EEC data that accounted for both charged and neutral final-state particles, provided statistical and systematic uncertainties or their quadrature sum, and had 50 bins or more across the entire $\chi$ range. 
We end up with 8 datasets from collaborations OPAL~\cite{OPAL:1991uui}, SLD~\cite{SLD:1994idb}, TOPAZ~\cite{TOPAZ:1989yod}, TASSO~\cite{TASSO:1987mcs}, MARKII~\cite{Wood:1987uf}, and MAC~\cite{Fernandez:1984db}, as summarized in Table~\ref{table1}. The fit range is constrained around the peak region of $145 \degree < \chi < 175 \degree$.

{\it \textbf{Fit Method.}} 
In this Letter, we fit the strong coupling constant $\as(m_Z)$, the NP Sudakov factor, and the NP CS kernel. Therefore, in total we have 4 fitting parameters:  $\as(m_Z)$, $a_1$, $a_2$, and $g_2$. To obtain optimal parameters, we apply the $\chi^2$ minimization method. For a given set of parameter values $\left\{\mathbf{p}\right\}$, the total $\chi^2\left(\left\{\mathbf{p}\right\}\right) = \sum_{i} \left(T_i\left(\left\{\mathbf{p}\right\}\right)-E_i\right)^2 / {\sigma_i^2}$. Here, we sum over all experimental data points; for the $i$-th data point, $E_i$ represents the measured EEC, and the uncertainty $\sigma_i$ equals the quadrature sum of its statistical and systematic uncertainties; $T_i\left(\left\{\mathbf{p}\right\}\right)$ represents our theoretical prediction for EEC with parameter values $\left\{\mathbf{p}\right\}$.

{\it \textbf{Treatment of Uncertainties.}}
We account for two sources of uncertainties in our analyses. The first stems from the fit of the strong coupling constant and the parameters of the NP Sudakov factor and the NP CS kernel. We refer to this as the fit uncertainty. To quantify it, we apply the standard replica method \cite{Bacchetta:2017gcc, Alrashed:2023xsv} using 200 replicas. For each replica, a random Gaussian noise---scaled by the experimental uncertainty---is added to the EEC data points. By fitting the parameters of each replica, we obtain 200 sets of optimal values. The median value for each parameter serves as the central result, while the upper and lower fit uncertainties correspond to the range that encompasses the central 68\% of the parameter values.

Secondly, we account for theoretical uncertainty, which arises from higher-order perturbative corrections in the hard and jet functions. To explore this uncertainty, we use the scan method \cite{Bell:2023dqs}. In this approach, the initial resummation scales, $\mu_H^i$, $\zeta_J^i$, and $\mu_J^i$, are varied from their canonical values by random factors within the range $[\frac{1}{2}, 2]$. When $\mu_H^i$ is shifted below its canonical value, we freeze it at 1 GeV for large $b$ to prevent entering the sub-GeV regime. We generate 200 sets of initial scales and perform fitting using the original experimental data, resulting in 200 sets of optimal parameters. The theoretical uncertainty for each parameter is then determined by the range of the central 68\% of these optimal parameter values.

\begin{table*}[htb]
\centering
\begin{tabular}{|c|c|c|c|c|c|}
\hline
& $\as(m_Z)$ & $a_1$~(GeV$^{a_2}$) & $a_2$ & $g_2 \ (\rm{GeV}^{2})$ & $g'_2$\\
\hline
& & & & & \\
\ Fit 1 \quad
& \ $0.1193^{+ 0.0009 + 0.0008}_{- 0.0009  - 0.0013}$ \quad
& \ $0.530^{+ 0.060 + 0.051}_{- 0.067 - 0.052}$ \quad
& \ $1.152^{+ 0.045 + 0.064}_{- 0.049 - 0.077}$ \quad
& \ $0.076^{+ 0.014 + 0.015}_{- 0.013 - 0.012}$ \quad
& N/A \\
& & & & & \\
\hline
& & & & & \\
\ Fit 2 \quad
& \ $0.1193^{+ 0.0010 + 0.0009}_{- 0.0008 - 0.0013}$ \quad
& \ $0.527^{+ 0.030 + 0.051}_{- 0.036 - 0.058}$ \quad 
& \ $1.152^{+ 0.040 + 0.065}_{- 0.038 - 0.079}$ \quad
& N/A
& \quad $0.194^{+ 0.019 + 0.037}_{- 0.020 - 0.028}$ \quad \\
& & & & & \\
\hline
\end{tabular}
\caption{Central parameter values along with fit and theory uncertainties corresponding to the $68\%$ probability around the median.}
\label{tab:Params}
\end{table*}

{\it \textbf{Numerical Results.}}
The central parameter values and uncertainties are detailed in Table~\ref{tab:Params}, and Fit 1's prediction is shown in Fig~\ref{FIG1}. As shown in Table.~\ref{table1}, both Fit 1 and Fit 2 achieved $\chi^2/{d.o.f} = 0.95$, demonstrating that our model accurately describes the EEC in the back-to-back region ($145^\circ<\chi<175^\circ$). Additionally, the model's predictions remain valid in the unfitted region ($115^\circ<\chi<145^\circ$).

Our extracted value of $\alpha_s(m_Z)$ aligns closely with the average value $0.1189 \pm 0.0037$ obtained from $e^+ e^-$ jet and event shape measurements, and is consistent with the Particle Data Group's world average of $0.1180 \pm 0.0009$ \cite{ParticleDataGroup:2024cfk} within uncertainty. With $a_2 = 1.152$, both fits' NP Sudakov factor are close to linear. 
We have also checked that fits performed at $\mathrm{N}^3\mathrm{LL}'$ yield central parameter values similar to those at $\mathrm{N}^3\mathrm{LL}$~\cite{Ebert:2020sfi}.
For consistent matching with the NLO non-singular contribution, we report only the $\mathrm{N}^3\mathrm{LL}$ results.

\begin{figure}
    \centering
    \includegraphics[width =\columnwidth]{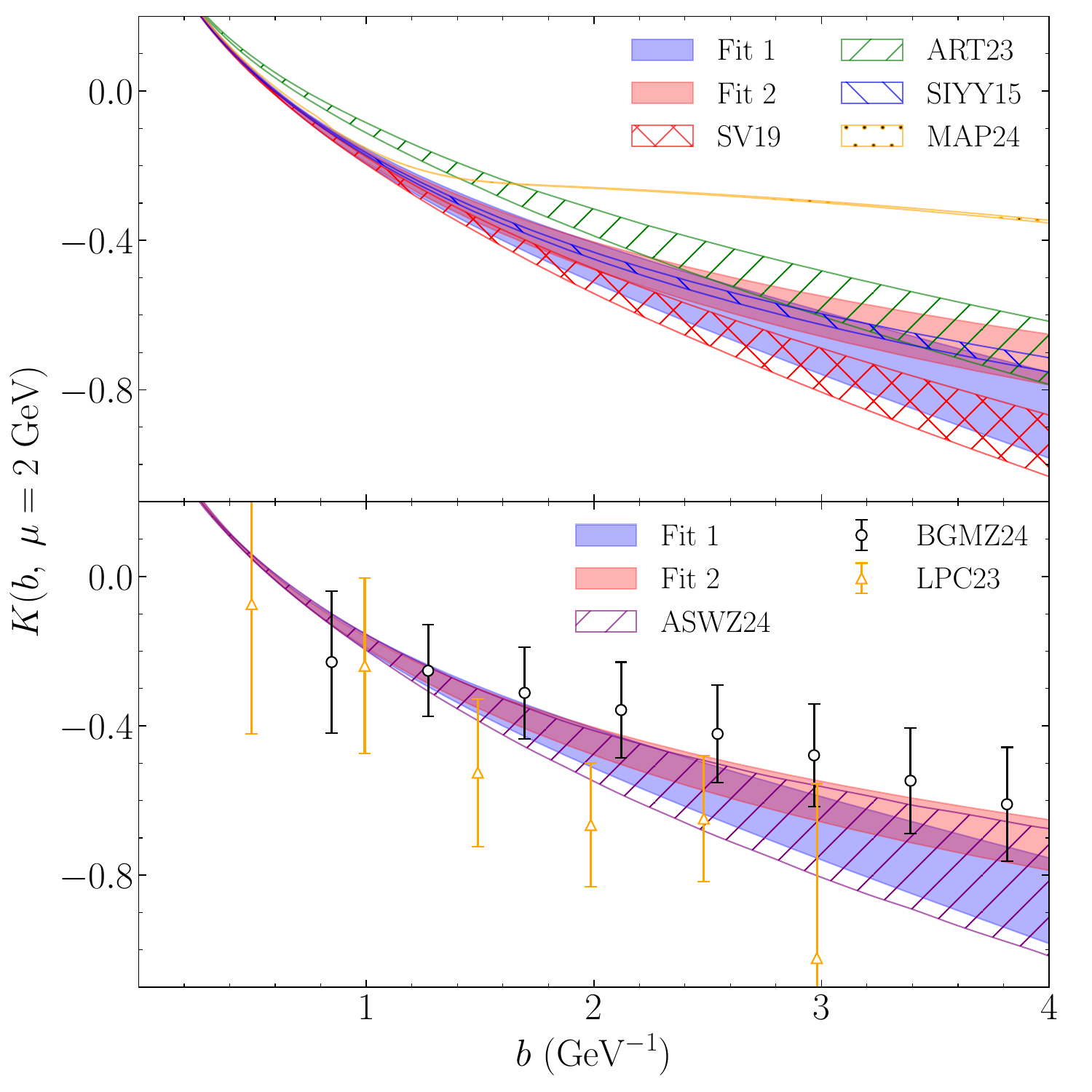} 
    \caption{We compare the full CS kernel $K(b, \mu=2 \rm{GeV})$ extracted from EEC (Fit 1 \& Fit 2) with other phenomenological extractions in SV19~\cite{Scimemi:2019cmh}, SIYY15~\cite{Sun:2014dqm}, ART23~\cite{Moos:2023yfa}, and MAP24~\cite{Bacchetta:2024qre} on the upper plot, and to lattice QCD calculations from ASWZ24~\cite{Avkhadiev:2024mgd}, BGMZ24~\cite{Bollweg:2024zet}, and LPC23~\cite{LatticePartonLPC:2023pdv} in the lower plot.}
    \label{CS}
\end{figure}

Fig.~\ref{CS} presents the full CS kernel $K$ extracted from EEC at $\mu = 2$ GeV, compared against phenomenological extractions~\cite{Scimemi:2019cmh, Sun:2014dqm, Moos:2023yfa, Bacchetta:2024qre} (upper panel) and lattice QCD results~\cite{Avkhadiev:2024mgd, Bacchetta:2024qre, Boussarie:2023izj} (lower panel). We display only the fit uncertainty bands to align with other studies. Notably, ART23~\cite{Moos:2023yfa} used solely Drell--Yan (DY) process data, while SV19~\cite{Scimemi:2019cmh}, SIYY15~\cite{Sun:2014dqm}, and MAP24~\cite{Avkhadiev:2024mgd} incorporated both semi-inclusive deep inelastic scattering (SIDIS) and DY. The parametrization of NP CS kernels for SV19 and SIYY15 correspond to those in our Fit 1 and Fit 2, and MAP24's simple quadratic parametrization $D_{\rm{NP}}^{\rm{MAP24}} = \tilde g_2 b^2$ can also be found in older extractions \cite{Bacchetta:2022awv,Bacchetta:2017gcc}. ART23's parametrization is $D_{\rm{NP}}^{\rm{ART23}} = b b_* (\tilde g_2 + \tilde g_3 \ln{(b_*/b_{\rm{max}})})$. SV19 and ART23 treat $b_{\rm{max}}$ as a fit parameter, while we set $b_{\rm{max}} = 2e^{-\gamma_E} \ \rm{GeV}^{-1}$. All three lattice QCD results for the CS kernel were based on quasi TMD wave functions. BGMZ24~\cite{Bollweg:2024zet} and LPC23~\cite{LatticePartonLPC:2023pdv} reported their CS kernels at discrete $b$ values, whereas ASWZ24~\cite{Avkhadiev:2024mgd} provided a continuum extrapolation. 

Our two extractions of the CS kernel overlap within the 68 \% fit uncertainty for $b<\SI{4}{GeV^{-1}}$. In comparison with other phenomenological fits, Fit 1 is consistent with SV19, while the uncertainty band of Fit 2 entirely encompasses that of SIYY15. Despite having a different functional form, ART23 also shows reasonable agreement with Fit2. We note that MAP24 deviates from both our fits and other phenomenological extractions presented in the plot, as well as recent extractions in~\cite{Isaacson:2023iui,Aslan:2024nqg}. When compared to lattice results, the continuum extrapolation from ASWZ24 closely matches our extraction, covering most of the uncertainty bands for both Fit 1 and Fit~2. Additionally, results from BGMZ24 and LPC23 align with our extractions within the 68\% uncertainty level.

{\it \textbf{Conclusions.}} 
We have analyzed the EEC in $e^+ e^-$ collisions in the back-to-back region using the TMD factorization formalism at N$^3$LL accuracy and, for the first time, performed a simultaneous fit of the strong coupling constant $\alpha_s(m_Z)$ and the non-perturbative contribution to the Collins--Soper (CS) kernel. 
Traditionally, the CS kernel has been derived from lattice QCD or phenomenological fits to SIDIS and DY process cross sections. Our method, based on the EEC, offers a cleaner extraction by avoiding the complexities of non-perturbative TMD hadron structures present in conventional SIDIS and DY extractions.

These fits, using two models for the non-perturbative CS kernel, yield strong coupling constants consistent with the world average. Despite slight model-specific dependencies, our CS kernel aligns with lattice QCD results and most other recent phenomenological extractions.
This remarkable alignment offers compelling evidence for the universality of the non-perturbative Collins--Soper kernel, and opens exciting opportunities for joint fits of the EEC experimental data alongside other TMD processes, such as SIDIS and DY production, which are highly promising to be explored at the Large Hadron Collider and the future Electron--Ion Collider.

{\it Acknowledgments.} 
This work is supported by the National Science Foundation under grant No.~PHY-1945471.
This work is also supported by the U.S. Department of Energy, Office of Science, Office of Nuclear Physics, within the framework of the Saturated Glue (SURGE) Topical Theory Collaboration. 
We thank the lattice groups of \cite{Avkhadiev:2024mgd, Bollweg:2024zet,LatticePartonLPC:2023pdv} for providing lattice data and the group of \cite{Bacchetta:2024qre} for providing the phenomenological band.

\bibliographystyle{JHEP-2modlong.bst}
\bibliography{refs}

\end{document}

%% file: packages.tex
\usepackage[utf8]{inputenc}

\usepackage{amsmath}
\usepackage{tikz}
\usepackage{siunitx}
\usepackage{physics}
\usepackage{slashed}

\usepackage{overpic}

\usepackage{graphicx}

\usepackage{mathtools}
\usepackage{physics}

\usepackage{xintfrac}

%% file: definitions.tex
\newcommand{\as}{\alpha_\text{s}}